\documentclass[12pt]{article}

\global\arraycolsep=1pt
\usepackage[dvips]{graphicx}
\usepackage{amsmath}
\usepackage{amssymb}

\newcommand{\beq}{\begin{equation}}
\newcommand{\eeq}{\end{equation}}
\newcommand{\beqa}{\begin{eqnarray}}
\newcommand{\eeqa}{\end{eqnarray}}

\renewcommand{\theequation}{\thesection.\arabic{equation}}
\renewcommand{\thefootnote}{\fnsymbol{footnote}}

\def\O{{\mathcal{O}}}

\begin{document}

\begin{titlepage}
\begin{flushright}
{
OCU-200\\
AP-GR-10
}
\end{flushright}
\vspace{0.5cm}
\begin{center}
{\Large \bf
Compact Einstein Spaces based on Quaternionic K\"ahler Manifolds
}%
\lineskip .75em
\vskip1.0cm
{\large Mitsuo Hiragane\footnote{e-mail: hiragane-m@rio.odn.ne.jp},
Yukinori Yasui\footnote{e-mail: yasui@sci.osaka-cu.ac.jp} and 
Hideki Ishihara\footnote{e-mail: ishihara@sci.osaka-cu.ac.jp}}
\vskip 1.0em
{\large\it Department of Physics, Osaka City University \\
Sumiyoshi-ku, Osaka, 558-8585, Japan}
\end{center}
\vskip0.5cm

\begin{abstract}


We investigate the Einstein equation with a positive cosmological 
constant for $4n+4$-dimensional 
metrics on  bundles over Quaternionic K\"ahler base manifolds 
whose fibers are 4-dimensional Bianchi IX manifolds. 
The Einstein equations are reduced to a set of non-linear 
ordinary differential equations. 
We numerically find inhomogeneous compact Einstein spaces 
with orbifold singularity.

\end{abstract}
\end{titlepage}

\renewcommand{\thefootnote}{\arabic{footnote}}
\setcounter{footnote}{0}

\section{Introduction}
\setcounter{equation}{0}
Compact Einstein spaces with positive-definite metrics have been 
studied extensively: 
first, the spaces are considered as candidates of compact internal spaces 
which are admitted in higher-dimensional gravitational theories\cite{Duff}; 
and secondly, the spaces are expected that they may dominate the path 
integrals of quantum gravity\cite{EQG}. 

There are many examples of homogeneous Einstein metrics
to be found 
in a lot of literature, but they are very exceptional among general 
Einstein spaces. 
In contrast, the inhomogeneous Einstein metrics would cover wide range, 
but our knowledge for them seems to be largely limited. 
The first example of inhomogeneous compact Einstein space 
which is a solution to the Einstein equation 
with a positive cosmological constant, 
\beqa
	\hat R_{ab} = \Lambda \hat g_{ab}, 
\label{Einstein_eq}
\eeqa
was constructed by Page\cite{rotating} by taking a limit of the Euclidean 
Kerr-de Sitter solution, and it was generalized by B\'erard-Bergery\cite{BB}.

In the absence of any general understanding of the solution to 
the Einstein equation \eqref{Einstein_eq}, one of the standard strategies 
for constructing 
inhomogeneous examples is to study the space with cohomogeneity one metric: 
the space admits a Lie group action by isometries whose orbits span the 
space with codimension one. 
It is considered that the space which foliates 
into a sequence of homogeneous subspaces of codimension one. 
The Einstein equation for such metric reduces to a set of non-linear 
second order ordinary differential equations\cite{W}.

It is possible to replace the homogeneous subspaces by spaces with 
bundle structure. This idea originates from the Kaluza-Klein construction.
The Einstein equation translates into a coupled system of 
equations involving the Ricci curvatures of the fibers and base space, 
as well as the curvature of the connection. 
When we have a suitable choice for the bundle space and connection, 
the Einstein equation also reduces to a set of the ordinary differential 
equations. 

In this paper, we construct compact inhomogeneous examples of 
the Einstein metric with a positive cosmological constant in 4n+4 
dimensions on the spaces with bundle structure. 
More precisely we consider the union of principal $SO(3)$-bundles $P$ over 
Quaternionic K\"ahler manifolds, so our spaces may be identified locally 
with $I\times P$ for some interval $I$ of the real line. 
This geometrical setting was studied in \cite{PP1} 
and they constructed a family of inhomogeneous compact Einstein spaces. 
Our construction is a generalization of their system\cite{PP1,PP2,GPP}. 

In this framework, it is possible to take 
three types of boundary conditions at two endpoints of $I$ for completeness, 
i.e., two types of bolt singularities associated with the Quaternionic 
K\"ahler manifold and with its twistor space respectively, 
and the nut singularity describing $S^{4n+3}$-collapsing. 
We find new compact Einstein spaces which have the two types of 
bolts, numerically. 
The new solutions together with the known solutions give a unified description 
of compact Einstein spaces with a positive cosmological constant.

The paper is organized as follows: 
Section 2 contains our Kaluza-Klein metric ansatz and calculations of 
the Riemannian curvature. 
In section 3 we derive the Einstein equation after a short review of 
Quaternionic K\"ahler manifolds. We also prove the diagonalizability of 
the metric using the technique for the 4-dimensional Bianchi IX metric. 
In section 4, we discuss the boundary conditions 
and give asymptotic solutions near the boundaries. 
In section 5, known exact solutions are listed. In section 6 we present new 
solutions obtained by numerical integrations. Section 7 is devoted to 
summary and discussion. 

\section{Metric Ansatz and Curvature Calculation} \label{ch:2}
\setcounter{equation}{0}

In this section, we shall consider metrics on 
($m+4$)-dimensional manifolds $\widehat{M}$. 
To make the analysis manageable, 
we assume the following geometrical condition for $\widehat{M}$.
Let $\pi:P \longrightarrow M$ be a principal $SO(3)$-bundle 
over an $m$-dimensional Riemannian manifold ($M,g_M$) and 
$\phi$ be an $SO(3)$-connection on $P$.
The connection $\phi$ locally takes the form
\begin{equation}
	\phi=s^{-1} A s+s^{-1} ds, \quad s \in SO(3) . 
\label{eq:connection}
\end{equation}
Here, $A$ is an $so(3)$-valued local 1-form on $M$ and 
$s^{-1} ds$ is considered as the Maurer-Cartan form.
Let $\phi^i$ be the components of $\phi$ 
for the standard basis \{$E_i$\} of $so(3)$ 
which satisfies the Lie bracket relations $[E_i,E_j]=\sum_{k=1}^3 \varepsilon_{ijk}E_k$.
By using the left-invariant 1-forms $\sigma^i$ defined by 
$s^{-1} ds = \sum_{i=1}^3 \sigma^i E_i$,
the equation \eqref{eq:connection} can be also written as
\begin{equation}
\phi^i=\sum_{j=1}^3 A^j \O_{ji} + \sigma^i
\end{equation}
with the adjoint representation
\begin{equation}
s^{-1} E_i s=\sum_{j=1}^3 \O_{ij} E_j~.
\end{equation}
In this setting we consider metrics 
on $\widehat{M}$ which is locally the product space $I \times P$, 
where $I$ denotes some interval of {\bf{R}}. 
Given a metric $b_{ij}$ on $SO(3)$,
the Kaluza-Klein metric takes the form
\begin{equation}
\hat{g}=dt^2+\sum_{i,j=1}^3 b_{ij}(t) \phi^i \phi^j + f(t)^2 g_M \, . \label{eq:ometric}
\end{equation}
We can show that the matrix $b_{ij}(t)$ is diagonalizable for all $t$ 
under a certain condition of the base space $M$~
(see a proposition in section 3).
Thus we write the metric as  
\begin{equation}
	\hat{g}=dt^2+a(t)^2 (\phi^1)^2+b(t)^2 (\phi^2)^2+c(t)^2 (\phi^3)^2 
			+ f(t)^2 g_M  . 
\label{eq:metric}
\end{equation} 
If we impose the condition $a=b=c$,
then the metric \eqref{eq:metric} has an isometry $SO(3)$. 
Otherwise it has no such symmetry because of the explicit dependence
on the group element $\O_{ij}$. 

Now, we choose an orthonormal basis 
$\hat{e}^a=\{ \hat{e}^0, \; \hat{e}^i, \; \hat{e}^{\alpha} \; ; \; i=1 \sim 3,\; \alpha=1\sim m \}$ for \eqref{eq:metric},
\begin{equation}
\hat{e}^0=dt,\quad \hat{e}^i=a_i \phi^i,\quad \hat{e}^{\alpha}=f e^{\alpha}~, \label{eq:o.n.b}
\end{equation}
where $(a_i)=(a,~b,~c)$ and $e^{\alpha}$ is an orthonormal basis for $g_M$.
Then, the spin connection $\hat{\omega}_{ab}$ 
defined by $d\hat{e}^a = - \hat{\omega}_{ab} \wedge \hat{e}^b$
is calculated as
\begin{equation}
\begin{array}{c}
\hat{\omega}_{0i}= - \cfrac{\dot{a}_i}{a_i} \,\hat{e}^i\, ,\quad
\hat{\omega}_{0\alpha}= - \cfrac{\dot{f}}{f} \,\hat{e}^{\alpha}\, ,\quad
\hat{\omega}_{ij}= \cfrac{1}{2} \,{\displaystyle \sum_k} \varepsilon_{ijk} \cfrac{a_k^2-a_i^2-a_j^2}{a_i a_j a_k} \,\hat{e}^k,\\[15pt]
\hat{\omega}_{i\alpha}= \cfrac{a_i}{2 f^2}~ {\displaystyle \sum_{k,\beta}} \O_{ki} F^k_{\alpha \beta} \hat{e}^{\beta}\, ,\quad
\hat{\omega}_{\alpha \beta}= \omega_{\alpha \beta} -  {\displaystyle \sum_{k,i}} \cfrac{a_i}{2 f^2} ~\O_{ki} F^k_{\alpha \beta} \hat{e}^i\, ,
\end{array}
\end{equation}
where $F=\sum_i F^i E_i$ is the curvature 2-form of $A$ 
and $\omega_{\alpha \beta}$ is the spin connection on $M$.
The curvature 2-form $\widehat{\Omega}_{ab}
=d\hat{\omega}_{ab}+\sum_c \hat{\omega}_{ac} \wedge \hat{\omega}_{cb}$
is given by the following equations:
\begin{eqnarray*}
	\widehat{\Omega}_{0i} 
		&=& - \cfrac{\ddot{a}_i}{a_i} \hat{e}^0 \wedge \hat{e}^i 
           + \cfrac{1}{2} \,\displaystyle{\sum_{j,k}} \varepsilon_{ijk} 
			\left[\cfrac{\dot{a}_i}{a_j a_k} + \cfrac{\dot{a}_j}{a_j} \,
			\cfrac{a_k^2-a_i^2-a_j^2}{a_i a_j a_k} \, \right]
			\hat{e}^j \wedge \hat{e}^k  \\ 
        && - \cfrac{a_i}{2 f^2} \left(\cfrac{\dot{a}_i}{a_i} 
		- \cfrac{\dot{f}}{f} \right) \displaystyle{\sum_{k,\alpha,\beta}} 
			\O_{ki} F^k_{\alpha \beta} \hat{e}^{\alpha} \wedge \hat{e}^{\beta},
	\cr
	\widehat{\Omega}_{0\alpha} 
		&=& - \cfrac{\ddot{f}}{f} \hat{e}^0 \wedge \hat{e}^{\alpha} 
           - \cfrac{a_i}{2 f^2} \left(\cfrac{\dot{a}_i}{a_i} 
			- \cfrac{\dot{f}}{f} \right) \displaystyle{\sum_{k,i,\beta}} 
			\O_{ki} F^k_{\alpha \beta} \hat{e}^{i} \wedge \hat{e}^{\beta} ,
	\cr
	\widehat{\Omega}_{i\alpha} 
		&=& - \displaystyle{\sum_{j,\beta}} \left[ 
			\cfrac{\dot{a_i} \dot{f}}{a_i f} ~\delta_{ij} 
			\delta_{\alpha \beta} + \displaystyle{\sum_{k,l}} 
			\cfrac{1}{4 f^2} \varepsilon_{ijl} 
			\cfrac{a^2_i + a^2_j -a^2_l}{a_i a_j} ~
			\O_{kl} F^k_{\alpha \beta} 
-\displaystyle{\sum_{k,l,\gamma}} 
			\cfrac{1}{4 f^4} a_i a_j \O_{ki} F^k_{\beta \gamma} 
			\O_{lj}F^l_{\alpha \gamma} \right] \hat{e}^j \wedge \hat{e}^{\beta}
 \\
      	&&+ \cfrac{1}{2}\left( \cfrac{\dot{a_i}}{f^2} - a_i \cfrac{\dot{f}}{f} \right) \displaystyle{\sum_{k,\beta}} \O_{ki} F^k_{\alpha \beta} \hat{e}^0 \wedge \hat{e}^{\beta} 
                       - \cfrac{a_i}{2 f^3} \displaystyle{\sum_{k,\beta,\gamma}} \O_{ki} \left( D_{\gamma} F^k_{\alpha \beta}\right) \hat{e}^{\beta} \wedge \hat{e}^{\gamma}\, ,
\cr
	\widehat{\Omega}_{\alpha \beta} 
		&=& \Omega_{\alpha \beta} - \cfrac{{\dot{f}}^2}{f^2} \hat{e}^{\alpha} \wedge \hat{e}^{\beta} 
           - \displaystyle{\sum_{i,k,l} \sum_{\gamma,\delta}} \cfrac{a_i^2}{4 f^4} \left( \O_{ki} F^k_{\alpha \beta} \O_{li} F^l_{\gamma \delta} + \O_{ki} F^k_{\alpha \gamma} \O_{li} F^l_{\beta \delta} \right) \hat{e}^{\gamma} \wedge \hat{e}^{\delta} \\
           &&+ \displaystyle{\sum_{i,k}} \displaystyle{\sum_{\gamma}} \cfrac{a_i}{2 f^3} \O_{ki} \left( D_{\gamma} F^k_{\alpha \beta} \right) \hat{e}^i \wedge \hat{e}^{\gamma} - \displaystyle{\sum_{i,k}} \left( \cfrac{\dot{a_i}}{f^2}- a_i \cfrac{\dot{f}}{f^3} \right) \O_{ki} F^k_{\alpha \beta} \hat{e}^0 \wedge \hat{e}^i \\
            &&+ \displaystyle{\sum_{i,j}} \displaystyle{\sum_{k,l}} \displaystyle{\sum_{\gamma}} \cfrac{a_i a_j}{4 f^4} \O_{ki} F^k_{\alpha \gamma} \O_{lj} F^l_{\gamma \beta} \hat{e}^i \wedge \hat{e}^j \\
		&&- \cfrac{1}{2 f^2} \displaystyle{\sum_k} F^k_{\alpha \beta} \left[ \O_{k1} \cfrac{a^2_2 + a^2_3 -a^2_1}{a_2 a_3} \; \hat{e}^2 \wedge \hat{e}^3 
	+ \O_{k2}\cfrac{a^2_3 + a^2_1 -a^2_2}{a_3 a_1}  \hat{e}^3 \wedge \hat{e}^1 
\right. \\
		&&\qquad \left.
	+ \O_{k3} \cfrac{a^2_1 + a^2_2 -a^2_3}{a_1 a_2} \hat{e}^1 \wedge \hat{e}^2 \right]\, ,
\end{eqnarray*}
where $\Omega_{\alpha \beta}$ is the curvature on $M$ 
and $D_{\gamma} F^k_{\alpha \beta}$ is the gauge covariant derivative of $F$,
\begin{equation}
\begin{aligned}
D_{\gamma} F^k_{\alpha \beta}= \nabla_{\gamma} F^k_{\alpha \beta} + \sum_{i,j} \varepsilon_{kij} A^i F^j_{\alpha \beta}~. 
\end{aligned}
\end{equation}
The remaining components $\widehat{\Omega}_{ij}$ ($i,j =1 \sim 3$) are 
\begin{equation}
\begin{aligned}
\widehat{\Omega}_{12} 
	=& \left[ \cfrac{\dot{a}_3}{a_1 a_2} - \cfrac{\dot{a}_1}{a_1} \; 
		\cfrac{a^2_3+a^2_1-a^2_2}{2 a_1 a_2 a_3} - \cfrac{\dot{a}_2}{a_2} \; 
		\cfrac{a^2_3-a^2_1+a^2_2}{2 a_1 a_2 a_3} \right] \hat{e}^0 
		\wedge \hat{e}^3 \\
	&- \left[ \cfrac{\dot{a}_1 \dot{a}_2}{a_1 a_2} + \cfrac{a^2_3-a^2_1-a^2_2}
		{2 a^2_1 a^2_2}+\cfrac{a^4_3 - (a^2_1 - a^2_2)^2}{4 a^2_1 a^2_2 a^2_3} 
		\right] \hat{e}^1 \wedge \hat{e}^2 \\
	&+ \cfrac{1}{4 f^2} \displaystyle{\sum_k} \sum_{\alpha,\beta}
		\left[ \cfrac{a^2_3 - a^2_1 - a^2_2}{a_1 a_2}~\O_{k3}F^k_{\alpha\beta} 
		- \cfrac{a_1 a_2}{f^2} ~\displaystyle{\sum_{l,\gamma}} 
		\O_{k1} F^k_{\alpha \gamma} \O_{l2} F^l_{\beta \gamma} \right] 
		\hat{e}_{\alpha} \wedge \hat{e}_{\beta}~,
\end{aligned}
\end{equation}
and $\widehat{\Omega}_{23},\; \widehat{\Omega}_{31}$ 
are given by cyclic permutations $1\rightarrow 2 \rightarrow 3 \rightarrow 1$. 
The non-zero components of Ricci tensor $\widehat{R}_{ab}$ become
\begin{equation}
\begin{aligned}
	\widehat{R}_{00} 
		&= - {\sum_i} \frac{\ddot{a_i}}{a_i} 
		- m \frac{\ddot{f}}{f} ~,\\[5pt]
	\widehat{R}_{ij} 
		&= \left[- \frac{d}{dt} \left( \frac{\dot{a_i}}{a_i} \right) 
		- \frac{\dot{a_i}}{a_i} \left( \sum_k 
		\frac{\dot{a_k}}{a_k} + m \frac{\dot{f}}{f} \right) \right] 
		\delta_{ij} 
		+ R^{SO(3)}_{ij} + \frac{a_i a_j}{4 f^4}~ \sum_{k,l}
		\sum_{\alpha,\gamma} \O_{ki} F^k_{\alpha \gamma} 
		\O_{lj} F^l_{\alpha \gamma}~,\\[10pt]
	\widehat{R}_{\alpha \beta} 
		&= \left( -\frac{\ddot{f}}{f} - (m-1) \frac{\dot{f}^2}{f^2} 
		- \sum_i \frac{\dot{a}_i \dot{f}}{a_i f} \right) 
		\delta_{\alpha \beta} 
		+\cfrac{1}{f^2} ~R_{\alpha \beta}- \sum_{i,k,l} 
		\sum_{\gamma} \frac{a^2_i}{2 f^4} ~\O_{ki} 
		F^k_{\alpha \gamma} \O_{li} F^l_{\beta \gamma}~,\\[7pt]
	\widehat{R}_{i \alpha} 
		&= \frac{a_i}{2f^3}~\sum_{k,\beta} \O_{ki} D_{\beta}F^k_{\alpha\beta}.
\end{aligned} 
\label{eq:ric}
\end{equation}
Here $R_{\alpha \beta}$ denotes the Ricci tensor on $M$
and $R^{SO(3)}_{ij}$ the Ricci tensor on $SO(3)$. Explicitly
the non-zero components of $R^{SO(3)}_{ij}$ are given by
\begin{equation}
R^{SO(3)}_{11} = \cfrac{a^4_1 - (a^2_2 - a^2_3)^2}{2 a^2_1 a^2_2 a^2_3}\, ,\;
R^{SO(3)}_{22} = \cfrac{a^4_2 - (a^2_3 - a^2_1)^2}{2 a^2_1 a^2_2 a^2_3}\, ,\;
R^{SO(3)}_{33} = \cfrac{a^4_3 - (a^2_1 - a^2_2)^2}{2 a^2_1 a^2_2 a^2_3}\, .
\end{equation}

\section{Einstein equation based on Quaternionic K\"ahler 
Manifold} \label{ch:3}
\setcounter{equation}{0}

In order to solve the Einstein equation,
we need a further assumption on the base space $M$ 
of the principal bundle $P$. Then we will obtain a generalization
of the ordinary differential equations studied 
in \cite{PP1,PP2,GPP}.

Let ($M,g_M$) be a 4n-dimensional Quaternionic K\"ahler manifold (QK manifold).
It has a set of three almost complex structures $J^a(a=1 \sim 3)$ 
which satisfy the quaternion algebra 
\begin{equation}
J^a J^b = - \delta_{ab} + \varepsilon_{abc} J^c \, , \label{eq:qk1}
\end{equation}
and one can find local 1-forms $A^i$ such that
\begin{equation}
\nabla J^a=\varepsilon_{abc}A^b \otimes J^c\, , \label{eq:qk2}
\end{equation} 
where $\nabla$ denotes the Levi-Civita connection\cite{ishihara}.
According to \cite{GP}, we introduce 
an orthonormal basis with 2-indices \{$e^{\mu i}$\};
\begin{equation}
g_M=\sum_{\mu=0}^3 \sum_{i=1}^{n} e^{\mu i} \otimes e^{\mu i}~.
\end{equation}
Then the 2-forms $J^a$ are given by 
\begin{equation}
\begin{aligned}
J^1=\sum_{i=1}^3 e^{0i} \wedge e^{1i} + e^{2i} \wedge e^{3i} ~,\\
J^2=\sum_{i=1}^3 e^{0i} \wedge e^{2i} + e^{3i} \wedge e^{1i} ~,\\
J^3=\sum_{i=1}^3 e^{0i} \wedge e^{3i} + e^{1i} \wedge e^{2i} ~.
\end{aligned}
\end{equation}
Furthermore, the $SO(3)$-connection $A=\sum_{i=1}^3 A^i E_i$ 
is $c_1$-self-dual in the sense of \cite{GP}, i.e., 
the Yang-Mills curvature $F=dA+A\wedge A$ satisfies the following 
self-dual equation
\begin{equation}
	\ast F=c_1 \; F \wedge \Omega^{n-1}, \quad 
	\Omega=\sum_{a=1}^3 J^a \wedge J^a  
\end{equation}
with $c_1=6n/(2n+1)!$ .
We note that the connection $A$ automatically satisfies the Yang-Mills equation
like a 4-dimenional ordinary instanton. In fact
\begin{equation}
D\ast F=c_{1}D(F \wedge \Omega^{n-1})=0~,
\end{equation}
where we have used $d\Omega=0$ by \eqref{eq:qk2} and the Bianchi 
identity $DF=0$. 
It is known that any QK manifold is Einstein and $F$ simply takes the form\cite{ishihara}\cite{Salamon} 
\begin{equation}
F^a=\cfrac{\lambda}{n+2}\; J^a~, \label{eq:qk4}
\end{equation}
where $\lambda$ is the Einstein constant for the metric $g_M$.

Now let us turn to the evaluation of the Ricci tensor \eqref{eq:ric}.
From QK geometry, we saw that the $SO(3)$-connection $A$ is $c_1$-self-dual 
and its curvature satisfies the quaternionic relations.
Thus, when we take a 4n-dimensional QK manifold as the base space
with self-dual connection, we have
\footnote{We have used a normalization 
$\lambda=n+2$ in \eqref{eq:qk3} and \eqref{eq:qk5} since we shall consider only $\lambda >0$ cases.} 
\begin{equation}
R_{\alpha \beta}=(n+2)~ \delta_{\alpha \beta}~, \quad \sum_{\alpha} D_{\alpha} F^i_{\alpha \beta}=0
\label{eq:qk3}
\end{equation}
and
\begin{equation}
\begin{aligned}
&\sum_{\alpha,\beta}F^i_{\alpha \beta}F^i_{\alpha \beta}=4n\delta_{ij} ~, \\
&\sum_{k,l} \sum_{\gamma} \O_{ki} F^k_{\alpha \gamma} \O_{li} F^l_{\beta \gamma}
                 =\sum_{k,l} \O_{ki} \O_{li}~(\delta^{kl} \delta_{\alpha \beta}
                +\varepsilon_{klj}F^j_{\alpha \beta})=\delta_{ii} ~ \delta_{\alpha \beta}~. 
\end{aligned} \label{eq:qk5}
\end{equation}
These equations finally lead us to the (4n+4)-dimensional Einstein equation 
with a cosmological constant $\Lambda$:
\begin{equation}
\begin{aligned}
\cfrac{\ddot{a}}{a} &+ \cfrac{\ddot{b}}{b} + \cfrac{\ddot{c}}{c} + 4 n \cfrac{\ddot{f}}{f}=-\Lambda  ~.\\
\cfrac{\ddot{a}}{a} &= - \cfrac{\dot{a}}{a} \left( \cfrac{\dot{b}}{b} + \cfrac{\dot{c}}{c} + 4 n \cfrac{\dot{f}}{f} \right) + \cfrac{a^4 - (b^2-c^2)^2}{2 a^2 b^2 c^2} + n \cfrac{a^2}{f^4} - \Lambda ~,\\
\cfrac{\ddot{b}}{b} &= - \cfrac{\dot{b}}{b} \left( \cfrac{\dot{c}}{c} + \cfrac{\dot{a}}{a} + 4 n \cfrac{\dot{f}}{f} \right) + \cfrac{b^4 - (c^2-a^2)^2}{2 a^2 b^2 c^2} + n \cfrac{b^2}{f^4} - \Lambda ~,\\
\cfrac{\ddot{c}}{c} &= - \cfrac{\dot{c}}{c} \left( \cfrac{\dot{a}}{a} + \cfrac{\dot{b}}{b} + 4 n \cfrac{\dot{f}}{f} \right) + \cfrac{c^4 - (a^2-b^2)^2}{2 a^2 b^2 c^2} + n \cfrac{c^2}{f^4} - \Lambda ~,\\
\cfrac{\ddot{f}}{f} &= - \cfrac{\dot{f}}{f} \left( \cfrac{\dot{a}}{a} + \cfrac{\dot{b}}{b} + \cfrac{\dot{c}}{c} + (4 n -1 ) \cfrac{\dot{f}}{f} \right) - \cfrac{a^2 + b^2 + c^2}{2 f^4} + \cfrac{n+2}{f^2} - \Lambda ~.
\end{aligned} \label{eq:ein}
\end{equation}  
Here, if we impose the conditions $a=b=c\neq f$ and $a \neq b=c=f~$, 
then \eqref{eq:ein} yields the equation 
studied in \cite{PP1} and \cite{PP2}, respectively (see section 5).
Also the singular reduction $a\equiv 0~$ and $b=c$
gives the $(4n+3)$-dimensional Einstein equation discussed in \cite{GPP}.
In case of the trivial bundle $P=M\times SO(3)$, i.e., $A=0$, 
the Ricci tensor \eqref{eq:ric} directly yields the Einstein equation 
without the assumption of QK manifolds. 
This equation is simply given by dropping the $f^{-4}$ terms and replacing 
the dimension $4n$ by arbitrary one in \eqref{eq:ein}.

It is worth noting that the $8$-dimensional Einstein equation 
with vanishing $\Lambda$ 
is special in the sense that the $Spin(7)$ holonomy condition 
leads to the following first-order equations\cite{Hitchin,CGLP5};
\begin{equation}
\begin{aligned}
	\cfrac{\dot{a}}{a}&=\cfrac{a^2-(b-c)^2}{2 a b c} - \cfrac{a}{f^2}, \\
	\cfrac{\dot{b}}{b}&=\cfrac{b^2-(c-a)^2}{2 a b c} - \cfrac{b}{f^2},\\
	\cfrac{\dot{c}}{c}&=\cfrac{c^2-(a-b)^2}{2 a b c} - \cfrac{c}{f^2}, \\
	\cfrac{\dot{f}}{f}&=\cfrac{a+b+c}{2 f^2}.
\end{aligned}   
\label{eq:spin7}
\end{equation}
We can verify by substituting \eqref{eq:spin7} into \eqref{eq:ein} with $n=1$ 
that the metric is indeed Ricci-flat, 
and the explicit solutions were constructed 
in \cite{GPP, CGLP5, BS, CGLP3}.  

Finally we prove the following as mentioned in section 2~:
\begin{flushleft}
{\bf{ Proposition}} \quad Let $(M,~g_M)$ be a 4n-dimensional QK manifold with
$c_1$-self-dual connection. Then the Kaluza-Klein metric \eqref{eq:ometric} 
for the Einstein space 
can be put in the diagonal form \eqref{eq:metric} for all $t$.
\end{flushleft}

\begin{flushleft}
{\it{Proof.}} \quad We calculate the Ricci tensor for the
metric \eqref{eq:ometric}, and the result is given in Appendix A. 
Note that the off-diagonal component $R_{0i}$ takes the same form 
as the 4-dimensional Bianchi IX type cosmological model. 
So this enables us in the usual way (see \cite{Bogo}, for example) to
diagonalize the metric for all $t$~. The argument is based on 
an important property
of the Einstein equation, namely invariance under the right action of $SO(3)$.
Indeed, using the corresponding transformation of the connection
\begin{equation}
	\phi \longrightarrow \tilde{\phi}=s_0^{-1} \phi~ s_{0}~, \quad 
	s_{0} \in SO(3)~,
\label{right^action}
\end{equation}
we can diagonalize the fiber metric $b_{ij}$ at an initial time $t=t_0$.
Then the Einstein equation $R_{0i}=0$ leads to $\dot b_{ij}=0$ for $i\neq j$ 
at $t=t_0$, which implies the diagonality of the solution for all time.

\end{flushleft}

\section{Boundary Condition}
\setcounter{equation} {0}
In this section, we discuss boundary conditions for the Einstein 
equation \eqref{eq:ein}. 
Let us assume the following compact conditions of $\widehat{M}\simeq I \times P$~:
\begin{enumerate}
\item $I$ is the closed interval $[ t_1,t_2 ]$ ,
\item QK manifolds have a positive scalar curvature, namely $\lambda>0$.
\end{enumerate}
Furthermore we require that the singularities at the boundaries 
$t_1$ and $t_2$ are resolved by bolts or nuts; 
there are three types of resolutions, 
nut $\leftrightarrow$ nut,~ bolt $\leftrightarrow$ nut and  
bolt $\leftrightarrow$ bolt (see Fig.1). 
This means that near the boundary $P$ is locally of the form
\begin{equation}
	P \longrightarrow S^k \times B^{\ell} \quad (k+\ell =4n+3)~,
\end{equation}
where the radius of round $k$-sphere $S^k$ tends to zero 
at the boundary 
and the $\ell$-dimensional manifold $B^{\ell}$ remains non-vanishing.
We find there are three choices of the manifold $B^{\ell}$
consistent with the Einstein equation:
(B1)~QK manifold $M$ ($\ell$=4n), (B2)~twistor space $Z$ of the 
QK manifold ($\ell$=4n+2), (B3)~empty.
In the case of (B1) or (B2) the singularity can be resolved by bolt, 
we call these singularities QK-bolt and T-bolt, 
and the case (B3) by nut.
We summarize these boundary conditions as follows~:
\begin{enumerate}
\item[(B1)] QK-bolt
\begin{equation}
	\hat{g}\rightarrow dt^2 + t^2((\phi^1)^2+(\phi^2)^2+(\phi^3)^2)/4 
		+\alpha^2 g_M~,\quad t \rightarrow 0~. \label{eq:b1}
\end{equation}
$g_M$ denotes a metric on a QK manifold $M$.
\item[(B2)] T-bolt
\begin{equation}
	\hat{g}\rightarrow dt^2 + k^2 t^2(\phi^1)^2+\beta^2 g_Z~,\quad 
		t \rightarrow 0~. \label{eq:b2}
\end{equation}
$g_Z$ denotes a metric on a twistor space $Z$~,
\begin{equation}
	g_Z = (\phi^2)^2+(\phi^3)^2+\alpha^2 g_M~, 
\end{equation}
which shows $Z$ is an $S^2$-bundle over $M$, and the metric is 
K\"ahler-Einstein for $\alpha=1$. 
\item[(B3)] nut
\begin{equation}
	\hat{g}\rightarrow dt^2 + t^2((\phi^1)^2+(\phi^2)^2+(\phi^3)^2+g_M)/4,
	\quad t \rightarrow 0~. \label{eq:b3}
\end{equation}
\end{enumerate}
Here we have translated the boundary $t=t_1$ or $t_2$  to the origin $t=0$, 
and 
$\alpha~$, $\beta$ and $k$ are free parameters.
In the case (B3) the QK manifold is required to be {\bf{H}}$P(n)$, 
otherwise it yields the curvature singularity at the boundary $t=0~$ 
since it does not describe the $S^{4n+3}$-collapsing (nut singularity).
On the other hand, in the cases (B1) and (B2) 
the manifold $\widehat{M}$ would have an orbifold singularity at $t=0~$ 
though arbitrary QK manifolds are allowed.
Indeed, $\sum_i (\phi^i)^2/4$ represents the metric on $SO(3) \simeq S^3/Z_2~$ 
rather than $SU(2)\simeq S^3$ , 
and the range of the angle $\Theta=k \psi$ ($\phi^1=d\psi$ for the fixed twistor space coordinate) 
does not mean $0\leq \Theta < 2 \pi$ generally. For the case (B1),
if we choose the QK manifold {\bf{H}}$P(n)$~, then 
the orbifold singularity disappears
by lifting to an $SU(2)$-bundle (see section 5)~.   

\begin{figure}[h]
  \begin{center}
    \includegraphics[height=12pc]
		{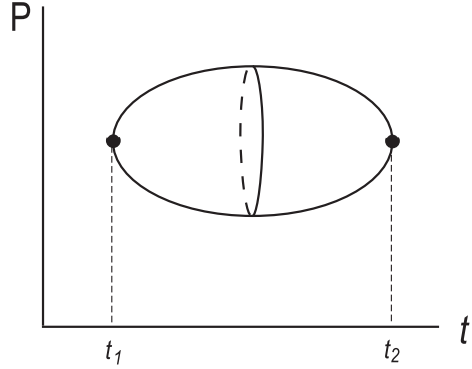}
\parbox{14cm}{  
\caption{\small Compact manifolds $\widehat{M}\simeq I \times P$. 
$I$ is $[t_1, t_2]$ and $P$ is a principal $SO(3)$-bundle 
over a Quaternionic K\"ahler manifold. Endpoints at $t_1$ and $t_2$ 
are boundaries which can take nut or bolts.
}}
  \end{center}
    \label{fig:bundle}
\end{figure}

Using the Einstein equation we find the following asymptotic behavior 
of the metric near the boundary: 
\begin{enumerate}
\item[(B1)] QK-bolt
\begin{equation}
\begin{aligned}
a&=t/2+a_3/6 ~t^3+\cdots ~, \\ 
b&=t/2+b_3/6 ~t^3+\cdots ~, \\ 
c&=t/2+c_3/6 ~t^3+\cdots ~, \\
f&=f_0+f_2/2 ~t^2+\cdots ~. \\
\end{aligned}
\label{asympt-QK}
\end{equation}
Here three of $f_0~,~a_3~,~b_3$ and $c_3$ are free parameters 
which satisfy the relation
\begin{equation}
\begin{aligned}
a_3+b_3&+c_3+ \cfrac{n(2+n)}{2 f_0^2} +\frac{1}{2}(1-n)\Lambda=0~, \\
\end{aligned}
\end{equation}
and the remaining coefficients are determined by them.

\item[(B2)] T-bolt
\begin{equation}
\begin{aligned}
	a&=k t + a_3/6 ~t^3+\cdots~, \\
	b&=c=b_0+b_2/2 ~t^2+\cdots~, \\
	f&=f_0+f_2/2 ~t^2+\cdots~, \\
\label{ex1}
\end{aligned}
\end{equation}
where $k\,,\,b_0\,,\,f_0$ are free parameters and the remaining 
coefficients are determined by them.
The cases $k=2$ and $1/N$ for a positive integer $N$ are exceptional and 
the expansion above must be modified by the following equations~:

1.~~$k=2$
\begin{equation}
\begin{aligned}
	a&=2~t+a_3/6 ~t^3+\cdots~, \\
	b&=b_0+b_1 ~t +b_2/2 ~t^2+\cdots~, \\
	c&=b_0-b_1 ~t +b_2/2 ~t^2+\cdots~, \\
	f&=f_0+f_2/2 ~t^2 +\cdots~,
\label{ex2}
\end{aligned}
\end{equation}
where $b_0\,,\,b_1\,,\,f_0$ are free parameters and remaining coefficients are determined by these.

2.~~$k=1/N$

\begin{equation}
\begin{aligned}
	a&=t/N+a_3/6 ~t^3+\cdots~, \\
	b&=b_0 +b_2/2 ~t^2+\cdots~, \\
	c&=\sum_{j=0}^{N-1} \frac{b_{2j}}{(2j)!}~t^{2j}
		+\frac{c_{2N}}{(2N)!}~t^{2N}+\cdots~, \\
	f&=f_0+f_2/2 ~t^2 +\cdots~,
\label{ex3}
\end{aligned}
\end{equation}
where $b_0\,,\,f_0\,,\,\delta_N=b_{2N}-c_{2N}$ are free parameters and remaining coefficients are determined by these. Derivation of \eqref{ex3} is viewed in 
Appendix B.

\item[(B3)] nut
\begin{equation} 
\begin{aligned}
a&=t/2+a_3/6 ~t^3+\cdots~, \\ 
b&=t/2+b_3/6 ~t^3+\cdots~, \\
c&=t/2+c_3/6 ~t^3+\cdots~, \\
f&=t/2+f_3/6 ~t^3+\cdots~, \\
\end{aligned}
\end{equation}
where $a_3,~b_3,~c_3$ and $f_3$ are parameters 
satisfying $a_3+b_3+c_3+4n f_3+\Lambda/2=0~$.
\end{enumerate}

\section{Example}
\setcounter{equation}{0}
The quaternionic projective space {\bf{H}}$P(n)$ is a typical example of 
QK manifolds. 
The Hopf fibration $S^{4n+3} \longrightarrow {\mathbf{H}}P(n)$ 
is a principal $SU(2)$-bundle over {\bf{H}}$P(n)$. It has a natural connection 
such that its horizontal space is the orthogonal complement to 
the fiber with respect to the standard metric on $S^{4n+3}$. 
This connection is $c_1$-self-dual\cite{Naga} and in case of $n=1$ 
the connection is the well-known BPS instanton.

We explicitly calculate the Kaluza-Klein metric based on 
${\mathbf{H}}P(n)$ according to sections \ref{ch:2} and \ref{ch:3}.
For the base space $M$={\bf{H}}$P(n)$, the standard metric is written as 
\begin{equation}
	g_{M}=\frac{4 d\bar{x}^{A}dx^A}{1+\bar{x}^{C}x^{C}}-
\frac{4\bar{x}^{A}dx^{A}\bar{x}^{B}dx^{B}}{(1+\bar{x}^{C}x^{C})^2}, 
\label{eq:s4}
\end{equation}
where $x^{A}=x_0^A+x_1^{A}i+x_2^{A}j+x_3^{A}k~~(A=1\sim n)$ are quaternionic coordinates
and $\bar{x}^{A}$ are their conjugates. Then the $c_1$-self-dual connection takes the form
\begin{equation}
A^{1}i+A^{2}j+A^{3}k=\frac{\bar{x}^{A}dx^{A}-d\bar{x}^{A}x^{A}}{1+\bar{x}^{B}x^{B}}~.
\end{equation}
Thus we have a Kaluza-Klein metric \eqref{eq:metric} on $I \times P$ explicitly,
and the Einstein equation is given by \eqref{eq:ein}. 

By the construction $P$ is  an $SO(3)$-bundle over {\bf{H}}$P(n)$.
In general, there is an obstruction to lifting an $SO(3)$-bundle 
over QK manifold to an $SU(2)$-bundle, 
i.e., the Marchiafava-Romani class $\varepsilon$.
A result of \cite{Salamon} says that $\varepsilon=0$ if and only if 
the QK manifold is {\bf{H}}$P(n)$, and so
in this case the total space $P$ lifts to the covering space $\tilde{P}$ 
as an $SU(2)$-bundle.
Actually $\tilde{P}$ is the total space of the Hopf fibration, $S^{4n+3} \simeq Sp(n+1)/Sp(n)$.

For completeness we present a summary of the known compact complete metrics 
based on the Hopf fibration. All these metrics are given 
as solutions to appropriate reductions of our equation.
\begin{enumerate}
\item[(R1)] reduction to one variable~;~$a=b=c=f$ \\ 
\begin{equation}
a(t)=1/2 ~\cos t ~, \quad -\pi/2 \leq t \leq \pi/2
\end{equation}
with $\Lambda =4n+3~$. This solution, which obeys the boundary condition 
\eqref{eq:b3} at $t=\pm \pi/2~$, 
represents the standard metric on $\widehat{M}=S^{4n+4}~$.
\item[(R2)] reduction to two variables~;~$a=b=c$ and $f$ \\
An explicit solution is given by
\begin{equation}
a(t)=1/2 ~\sin t \cos t~, \quad f(t)=1/2 ~\cos t~, \quad 0\leq t \leq \pi/2 
\end{equation}
with $\Lambda =4n+12~$. This solution, which obeys 
\eqref{eq:b1},~\eqref{eq:b3} at $t=0, ~\pi/2$, respectively,   
gives the standard metric on ${\mathbf{H}}P(n+1)~$.
Another solution satisfying \eqref{eq:b1} at the both endpoints was constructed numerically. 
It gives a metric on an $S^4$-bundle 
over ${\mathbf{H}}P(n)$, i.e., $\widehat{M}=
{\mathbf{H}}P(n+1) ~\sharp ~\overline{{\mathbf{H}}P(n+1)}$\cite{PP1}.
The case $n=1$ has been analytically established 
by B$\ddot{\mathrm{o}}$hm\cite{bohm}. 
\item[($\widetilde{\mathrm{R2}}$)] reduction to two variables~;~$a$ and $b=c=f$ \\
The situation is very parallel to (R2), 
but the topology is different as we will see shortly.
The Fubini-Study metric
 on {\bf{C}}$P(2n+2)$ 
belongs to this class. 
Explicitly, the metric is written as
\begin{equation}
a(t)=1/2 ~\sin t \cos t~, \quad b(t)=1/2~ \cos t~, \quad 0\leq t \leq \pi/2 
\end{equation}
with $\Lambda =4n+6~$. This solution obeys 
the boundary conditions \eqref{eq:b2},~\eqref{eq:b3} at $t=0~,\pi/2~$, 
respectively.
In this case the general solution was constructed by Page-Pope\cite{PP2} as
\begin{equation}
\hat{g}=\frac{(1-r^2)^{2n+1}}{P(r)}dr^2+\frac{4m_{1}^2P(r)}{(1-r^2)^{2n+1}}
(\phi^1)^2 +m_1(1-r^2)g_Z~, 
\end{equation}
where
\begin{equation}
	P=m_2 r+\frac{n+1}{m_1}Q_{2n+1}-\Lambda~Q_{2n+2}~, \quad 
	Q_n=\sum_{j=0}^{n} \binom{n}{j}\frac{(-r^2)^j}{1-2j}
\end{equation}
with integration constants $m_{i}$~(i=1,2).
Its special case satisfying \eqref{eq:b2} at the both endpoints gives 
a metric on an $S^2$-bundle over the twistor space $Z$, i.e., 
$\widehat{M}={\mathbf{C}}P(2n+2) ~\sharp ~\overline{{\mathbf{C}}P(2n+2)}$~. 
\end{enumerate}

\section{New Solutions}
\setcounter{equation}{0}

The solutions listed in the previous section have two 
of three possible boundaries at the endpoints: 
(B1) QK-bolt, (B2) T-bolt and (B3) nut. 
Correspondence between the solutions and boundaries is 
schematically shown in Fig.2. 
No solution is found which connects QK-bolt and 
T-bolt though the existence is naturally expected 
from the Fig. 2. 
The solutions with QK-bolt are found in the ansatz (R2), 
and the solutions with T-bolt are in the other ansatz 
($\widetilde{\mathrm{R2}}$), 
then we 
search new solutions which connect QK-bolt and T-bolt 
under a generalized ansatz 
in which the metric has three unknown variables. 

\begin{figure}[h]
  \begin{center}
    \includegraphics[height=18pc]
		{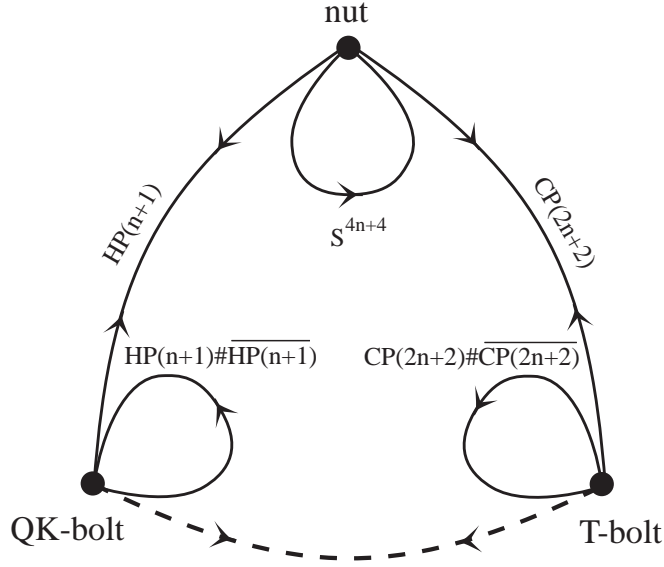}
\parbox{14cm}{
  \caption{\small Relation between boundaries and solutions.
}}
  \end{center}
    \label{fig:triangle}
\end{figure}

We assume $b=c$, hereafter, and treat $a, b, f$ as unknown metric functions 
of $t$. From \eqref{eq:ein} we obtain 
the evolution equations 
\begin{equation}
\begin{aligned}
  &\frac{\ddot a}{a}
	= - \frac{\dot a}{a}\left( 2\frac{\dot b}{b} 
			+  4n \frac{\dot f}{f} \right)  
 		+ \frac{{a}^2}{2\,{b}^4} 
		+ n \frac{{a}^2}{{f}^4} -\Lambda,
\\
  &\frac{\ddot b}{b} 
	= -  \frac{\dot b}{b} \left( \frac{\dot a}{a} + 
       \frac{\dot b}{b} + 4n \frac{\dot f}{f} \right) 
		- \frac{a^2-2b^2}{2\,{b}^4}  
		+ n \frac{{b}^2}{{f}^4} - \Lambda,
\\
  &\frac{\ddot f}{f}
	= -  \frac{\dot f}{f} \left( \frac{\dot a}{a}  
		+ 2\frac{\dot b}{b} + (4n-1)\frac{\dot f}{f}\right)  
		- \frac{a^2+2b^2}{2{f}^4} 
		+ \frac{n+2}{{f}^2} - \Lambda.
\end{aligned}
\label{dyn-eq}
\end{equation}
Taking a combination of \eqref{eq:ein} we also get the Hamiltonian constraint
\beqa
	&&2\frac{\dot b}{b}\left( 2\frac{\dot a }{a} + \frac{\dot b}{b} \right)  
	+ 4n\frac{\dot f}{f} \left( 2\frac{\dot a}{a} + 
       4\frac{\dot b}{b} + (4n-1) \frac{\dot f}{f}\right) \cr
&&
	\hspace{1cm} 
 	+ \frac{{a}^2 - 4{b}^2}{2{b}^4} 
	+ n \frac{a^2 + 2b^2}{f^4} 
	- 4n(n+2)\frac{1}{f^2} 
	 + (4n+2) \Lambda =0 .
\label{H-constraint}
\eeqa

We should integrate the equations \eqref{dyn-eq} 
from QK-bolt at $t=t_1=0$ 
as an \lq initial\rq\ value problem in $t$. 
Since $a$ and $b$ are zero at $t=0$, the set of ordinary differential 
equations \eqref{dyn-eq} is singular there. 
To deal with the singularity, we use the asymptotic form of the 
solutions \eqref{asympt-QK} near the singularity, and start numerical 
integrations at an initial time $t=t_i=0+\epsilon$, where $\epsilon$ is 
a small amount of time duration.

Equation \eqref{asympt-QK} gives the asymptotic 
behavior of $a, b$ and $f$ near QK-bolt:
\begin{equation}
\begin{aligned}
	a&=t/2+a_3/6 ~t^3+\cdots ~, \cr
	b&=t/2+b_3/6 ~t^3+\cdots ~, \cr
	f&=f_0+f_2/2 ~t^2+\cdots ~,
\end{aligned}
\label{asympt-B1}
\end{equation}
where  $f_0,~a_3$ and $b_3$ satisfy
\beqa
	a_3+2b_3+ \cfrac{n(2+n)}{2 f_0^2} +\frac{1}{2}\left(1-n \right)\Lambda=0~,
\label{const-cond-B1}
\eeqa
and $f_2$ is given by
\beqa
	f_2=\frac{1}{4} \left(\frac{n+2}{f_0}-\Lambda f_0\right).
\label{f2}
\eeqa
Introducing a parameter $g_0$ by
\beq
	g_0 = \frac{1}{3}(b_3-a_3),
\eeq
we can solve (\ref{const-cond-B1}) as
\begin{equation}
\begin{aligned}
	&a_3=\frac{1}{6}(n-1)\Lambda 
		- \frac{n}{6}(n+2)\frac{1}{f_0^2}-2g_0, \cr
	&b_3=\frac{1}{6}(n-1)\Lambda 
		- \frac{n}{6}(n+2)\frac{1}{f_0^2}+g_0.
\end{aligned}
\label{param-B1}
\end{equation}
Therefore, 
the initial condition at $t=t_i$ is given by 
\eqref{asympt-B1} \eqref{f2} and \eqref{param-B1}
with $t=\epsilon$.
We have two free parameters $f_0$ and $g_0$ to specify the 
initial condition near QK-bolt. 

We are interested in finding solutions for which $b$ and $f$ stay positive 
and finite, and $a$ returns to zero at some moment $t=t_2$. 
At this point the set of equations \eqref{dyn-eq} 
becomes singular again. 
We assume that this singularity is resolved 
by T-bolt \eqref{eq:b2}. 
If $k\neq 1$ the cone-type singularity appears at $t=t_2$ 
so we set $k=1$ for regularity. Then the condition \eqref{ex3} 
with $b=c$ ($\delta_1=0$) 
shows that the asymptotic behavior of $a, b$ and $f$ near $t=t_2$ 
should be
\beqa
	a&=& (t_2-t) + \bar a_3/6 ~(t_2-t)^3+\cdots~, 
\label{b2-a}\\
	b&=&\bar b_0+\bar b_2/2 ~(t_2-t)^2+\cdots~, 
\label{b2-b}\\
	f&=&\bar f_0+\bar f_2/2 ~(t_2-t)^2+\cdots,
\label{b2-f}
\eeqa
where $\bar b_0$ and $\bar f_0$ are free constants and 
$\bar a_3, \bar b_2$ and $\bar f_2$ are described 
by them as
\begin{equation}
\begin{aligned}
	\bar a_3&=-\frac{1}{\bar b_0^2}
		+n\frac{\bar b_0^2}{\bar f_0^4} 
		-\frac{2n(n+2)}{\bar f_0^2} 
		+2n\Lambda~, \\
	\bar b_2&=\frac{\bar b_0}{2} ~\left(\frac{1}{\bar b_0^2} 
		+ n \frac{\bar b_0^2}{\bar f_0^4} - \Lambda\right)~, \\
	\bar f_2&=\frac{\bar f_0}{2} ~\left(-\frac{\bar b_0^2}{\bar f_0^4}
		+\frac{2+n}{\bar f_0^2} -\Lambda\right)~. \\
\end{aligned}
\label{b2-higher}
\end{equation}
We stop numerical integrations at $t_f=t_2-\bar\epsilon$ 
($\bar\epsilon$ is a small constant) and check 
whether the values of 
$a, b, f$ and $\dot a, \dot b, \dot f$ agree with \eqref{b2-a}-\eqref{b2-f}.

Here, we define 
\beq
\begin{aligned}
	&V^1=\dot a(t_f)-\dot a_{T},\\
	&V^2=\dot b(t_f)-\dot b_{T},  
\end{aligned}
\eeq
where 
\begin{equation}
	\dot a_T = -1 - \bar a_3/2 ~\bar\epsilon^2, \quad
	\dot b_T = -\bar b_2 ~\bar\epsilon.
\label{cond-tf}
\end{equation}
A solution for \eqref{dyn-eq} defined on the region $[t_i, t_f]$ gives 
a map $\{(f_0, g_0)\} \rightarrow \{(V^1,V^2)\}$. 
Then we can regard $V=(V^1,V^2)$ as a vector field on a 2-dimensional 
plane parameterized by $(f_0, g_0)$. 
If $V=0$, $a$ and $b$ have the forms of \eqref{b2-a} and  
\eqref{b2-b} near $t=t_2$, 
in addition, it is shown from \eqref{H-constraint} $f$ should have the form of 
\eqref{b2-f} automatically.  
Then, vanishing of $V$ means the endpoint at $t=t_2$ is T-bolt.

Numerical integrations are done by using a fourth-order Runge-Kutta 
routine. 
We have verified that the constraint \eqref{H-constraint}, which is 
preserved by the evolution equations \eqref{dyn-eq}, 
holds in high accuracy in our numerical integrations. 
We also reproduce known solutions listed in the previous section 
when we set $a=b$ or $b=f$.

\begin{figure}[ht]
  \begin{center}
    \includegraphics[height=18pc]
		{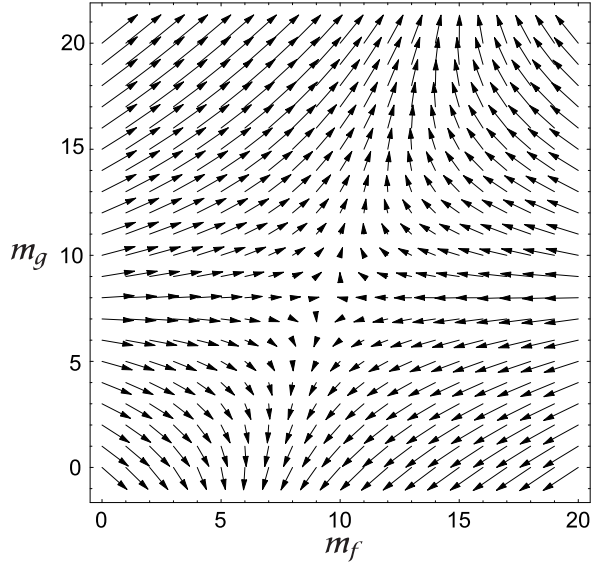}
    \label{arrows}
\parbox{14cm}{  
  \caption{\small Vector field $V$ on $(f_0,g_0)$-plane for $n=1$ case.
Parameters $m_f$ and $m_g$ give $f_0$ and $g_0$ as
$f_0 = 1.75333405 + 1.0\times 10^{-8}\times m_f$ and 
$g_0 = 0.002532820 + 1.0\times 10^{-9}\times m_g$. 
Horizontal and vertical components of the arrows are $V^1$ and $V^2$, 
respectively. The vector field vanishes on a point 
$(f_0^*, g_0^*) \sim (1.75334014, 0.002532828)$.
}}
  \end{center}
\end{figure}

Fig.3 shows the vector field $V$ on $(f_0, g_0)$-plane for $n=1$ 
case. 
We normalize $\Lambda=1$. 
There is a critical point $(f_0^*, g_0^*)$ on which 
$V$ vanishes. The map $(f_0^*, g_0^*) \mapsto V=0$ 
corresponds to the solution for \eqref{dyn-eq} 
which connects QK-bolt at $t=0$ and T-bolt at $t=t_2$. 
Evolution of $a, b$ and $f$ in $t$ is plotted 
in Fig.4. 

\begin{figure}[ht]
  \begin{center}
    \includegraphics[height=10pc]
		{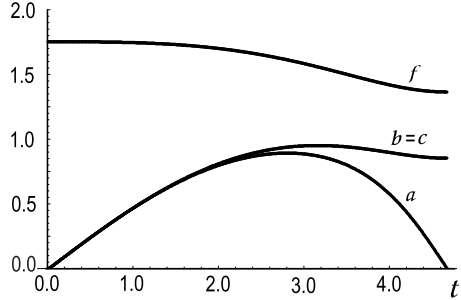}
\parbox{14cm}{  
  \caption{ \small
	Evolution of $a, b$ and $f$ in $t$ for the critical point $(f_0^*, g_0^*)$ 
in $n=1$ case.
}}
  \end{center}
   \label{evolution}
\end{figure}
\begin{table}[h]
	\begin{center}

 \begin{tabular}{rrrrrrrr}
\hline 
\multicolumn{1}{c}{$n$}& \multicolumn{1}{c}{$\sqrt{n+2}$} & 
 \multicolumn{1}{c}{$f_0^*$} & \multicolumn{1}{c}{$g_0^*$} & 
 \multicolumn{1}{c}{$\bar f_0$} & \multicolumn{1}{c}{$\bar b_0$}& 
 \multicolumn{1}{c}{$f_0^*-\bar f_0$} & \multicolumn{1}{c}{$t_2-t_1$}\\\hline
1 & 1.73205 & 1.75334 & 0.00253 & 1.36425 & 0.85289 & 0.38910 & 4.65747 \\ 
2 & 2.00000 & 2.05472 & 0.00441 & 1.74912 & 0.97181 & 0.30559 & 4.49746 \\ 
3 & 2.23607 & 2.30830 & 0.00633 & 2.03541 & 1.04715 & 0.27290 & 4.37347 \\ 
4 & 2.44949 & 2.52917 & 0.00811 & 2.27774 & 1.10000 & 0.25143 & 4.28367 \\ 
5 & 2.64575 & 2.72822 & 0.00968 & 2.49323 & 1.13927 & 0.23499 & 4.21713 \\ 
6 & 2.82843 & 2.91144 & 0.01107 & 2.68983 & 1.16969 & 0.22161 & 4.16625 \\ 
14 & 4.00000 & 4.07213 & 0.01767 & 3.91020 & 1.28394 & 0.16193 & 3.98370 \\ 
23 & 5.00000 & 5.06126 & 0.02091 & 4.93003 & 1.32869 & 0.13122 & 3.91686 \\ 
47 & 7.00000 & 7.04587 & 0.02420 & 6.95105 & 1.36954 & 0.09482 & 3.85836 \\ 
98 & 10.00000 & 10.03288 & 0.02616 & 9.96609 & 1.39204 & 0.06679 & 3.82719 \\ 
223 & 15.00000 & 15.02219 & 0.02726 & 14.97751 & 1.40431 & 0.04468 & 3.81052 \\ 
898 & 30.00000 & 30.01117 & 0.02795 & 29.98879 & 1.41173 & 0.02238 & 3.80052 \\ 
2498 & 50.00000 & 50.00671 & 0.02810 & 49.99328 & 1.41333 & 0.01344 & 3.79839 \\ 
9998 & 100.00000 & 100.00336 & 0.02816 & 99.99664 & 1.41399 & 0.00672 & 3.79749 \\
999998 & 1000.00000 & 1000.00034 & 0.02818 & 999.99966 & 1.41421 & 0.00067 & 3.79719 \\
\hline  
\end{tabular}

\parbox{14cm}{  
\caption[]{\small 
	Numerical parameters for $(4n+4)$-dimensional solutions.
Parameters $f_0^*, g_0^*$ for compact solutions, and 
parameters $\bar f_0, \bar b_0$, 
the difference $f_0^*-\bar f_0$, 
and the interval $t_2-t_1$ are listed. 
The cosmological constant is normalized as unity.
} 
}
	\end{center}
\label{}
\end{table}%


The vectors $V$ in Fig.3 turn $2\pi$ in the direction 
along a circle 
around the critical point $(f_0^*, g_0^*)$. 
Since the right hand sides of \eqref{dyn-eq} are regular 
in $t \in [t_i, t_f]$, the components of $V$ are  continuous functions 
with respect to $(f_0, g_0)$. 
Then there definitely exists a critical point on which 
$V=0$ inside the circle. 
Therefore this figure strongly suggests the existence of 
the solution with both QK-bolt and T-bolt.


As commented in section 5, the numerical metric with $k=1$ extends over
the QK manifold $M$ with an orbifold singularity at $t=0$, and extends smoothly over the twistor space $Z$ at the other endpoint $t=t_2$.
If we allow the cone-like singularity at T-bolt, $k$ can take an arbitrary 
value. The case is commented shortly in Appendix B.


In table 1 we list numerical values $f_0^*$ and $g_0^*$ 
for compact solutions of various dimensions $4n+4$ of the manifold. 
As $n$ is increased, $f(t)$ stays almost constant around $\sqrt{n+2}$ 
over the range of $t$, and the value of $\bar b_0$ at $t_2$ 
and the interval of the numerical solutions $t_2-t_1$ converges. 
Though the scale of the base space $f$ diverges in the order of $\sqrt{n}$ 
it seems that the 4-dimensional fiber metric converges. 
It should be noted that during the evolutions $\dot f/f$ is very small for 
large $n$ but the terms $4n \dot f/f$ in \eqref{dyn-eq} 
contribute in the same order of $\dot a/a$ and $\dot b/b$.

\section{Summary and Discussion}
\setcounter{equation}{0}
In the present work, we have made an investigation of higher dimensional 
compact Einstein manifolds. We can view such manifolds as the union
of principal $SO(3)$-bundles over Quaternionic K\"ahler manifolds. 
Globally, our compact manifold $\widehat{M}$ is considered as a fiber bundle
associated with a principal $G$-bundle $P$, i.e., $\widehat{M}=P \times_{G} F$~~($G=SO(3)$ or $SU(2)$)~. The fiber $F$ is a 4-dimensional manifold (orbifold)
with the Bianchi IX metric on which $G$ acts with cohomogeneity one, and
the base space $M$ is the Quaternionic K\"ahler manifold. 

Total space $\widehat{M}$ can be regarded as an evolution of $P$ 
over a finite \lq time\rq\ segment $[t_1, t_2]$. 
If we require the compactness, 
singularities at the boundaries $t_1$ and $t_2$ should be resolved by nut, 
Quaternionic K\"ahler(QK)-bolt or Twistor(T)-bolt. 
We found new solutions numerically 
which connect QK-bolt and T-bolt 
in the present paper, where T-bolt is characterized by a number $k$. 
To the extent of the three unknown variables $a, b=c, f$ for 
the metric form \eqref{eq:metric}, 
we could complete Fig.2 by using the new solutions 
(broken line) and already known solutions (solid lines).

Since $\widehat{M}$ has the bundle structure with the fiber $F$ on 
the base space $M$ 
then the Euler number is factorized as
\begin{equation}
	\chi(\widehat{M})=\chi(M)\chi(F)~.
\end{equation}
By using the Gauss-Bonnet theorem,
$\chi(F)$ is calculated as
\begin{equation}
\begin{aligned}
	\chi(F) 
	&=\frac{1}{32 \pi^2}\int_{F}
		\varepsilon_{abcd}~\Omega_{ab}\wedge \Omega_{cd}
    &=N_G(1/2+2 k)~,
\label{euler}
\end{aligned}
\end{equation}
where $N_G=1$ if $G=SO(3)$ and $N_G=2$ if $G=SU(2)$.
The factors 1/2 and $2k$ in \eqref{euler} represent the contribution 
from QK and T-bolts, respectively.


Let us consider the large $n$ limit.
Though the scale of the base space $f$ diverges in the order of $\sqrt{n}$ 
the coefficients \eqref{param-B1} of the local expansion near QK-bolt become
\begin{equation}
 a_3=-2g_0^{*}-\Lambda/6~,\quad b_3=g_0^{*}-\Lambda/6
\label{infty}
\end{equation}
and the ones \eqref{b2-higher} near T-bolt become
\begin{equation}
 \bar a_3=-\frac{1}{\bar b_0^2},\quad b_2=\frac{1}{2\bar b_0}-\frac{\bar b_0}{2}\Lambda,
\end{equation}
which might imply that 
the fiber metric converges to the 4-dimensional biaxial Bianchi IX metric 
\begin{equation}
 g_{\infty}
	=dt^2+\tilde{a}(t)^2 \sigma_1^2+\tilde{b}(t)^2(\sigma_2^2+\sigma_3^2)
\label{fiber}
\end{equation}
in the limit $n \rightarrow \infty$. 

When the total space connects QK-bolt and T-bolt, the fiber space with the 
metric \eqref{fiber} connects nut and bolt in 4-dimensions. 
It is known that the 4-dimensional biaxial Bianchi IX Einstein metrics
connecting nut and bolt singularities is either the self-dual 
Taub-NUT-de Sitter metric or the self-dual Eguchi-Hanson-de Sitter metric, 
and their common boundary represents the Fubini-Study metric on 
${\mathbf{C}}P(2)$\cite{CGLP7}. 
A question is whether the fiber metric coincides with the 
Bianchi IX Einstein metric in the limit $n \rightarrow \infty$. 
Indeed the expansion \eqref{infty} locally reproduces the behavior of the 
self-dual Eguchi-Hanson-de Sitter metric near the nut singularity, 
but the global metric is different 
since the value $k=1$ at the other endpoint (bolt singularity) for the 
fiber metric is not allowed for the 4-dimensional solution. 

This situation can be compared with the metric 
on ${\mathbf{H}}P(n+1) ~\sharp ~\overline{{\mathbf{H}}P(n+1)}$ 
solved by Page and Pope\cite{PP1} in the ansatz (R1)~; in the large $n$ limit,
$f$ stays constant and the fiber metric gives the standard $S^4$ metric i.e., 
the full metric approaches the direct product Einstein metric. 
In contrast, the numerical metric presented here is not the direct product 
metric even in the large $n$ limit, where the existence of 
the gauge field $A^i$ is important.


Finally we discuss the relation to the $Spin(7)$ holonomy metrics. 
In the beautiful papers\cite{Hitchin1}\cite{Hitchin2} Hitchin constructed 
a family of 4-dimensional self-dual Einstein metrics with positive 
Ricci curvature in a triaxial Bianchi IX form 
parameterized by an integer $N$. 
These solutions connect the two bolt singularities characterized by 
$k=2$ and by $1/N$. 
The metrics are explicitly given by the solution to the Painlev\'e VI 
equation\cite{Tod} and approach 
the Atiyah-Hitchin hyperk\"ahler metric\cite{A-H} 
with $Sp(1)$ holonomy 
in the limit $N\rightarrow \infty$. 

In 8-dimensional case, the asymptotically locally conical (ALC) $Spin(7)$ 
metrics would be counterparts of 
the Atiyah-Hitchin metric\cite{CGLP5,CGLP8,KY2}. 
The local behavior \eqref{ex2} at T-bolt with $k=2$ 
reproduces the ALC $Spin(7)$ metric when one of the free parameters is 
adjusted suitably. 
Furthermore the expansion \eqref{ex3} of the fiber metric 
at the other T-bolt with $k=1/N$ 
has the same form as the Hitchin metrics. 
It is tempting to expect that there is a series of Einstein metrics 
with positive Ricci curvature in 8-dimensions which approach the 
ALC $Spin(7)$ metric in a suitable large $N$ limit. 
Therefore, it is interesting to consider solutions 
in the general metric form \eqref{eq:metric} 
which connect two T-bolt singularities, $k=2$ and $1/N$. 
We leave this issue for further research.


\vskip10mm

\begin{center}
{\bf Acknowledgements}
\end{center}

We would like to thank H.Kanno for helpful discussions.
This work is supported in part by the Grant-in-Aid
for Scientific Research No.14540275 and No. 14570073.


\newpage

\section*{Appendix A}
\renewcommand{\theequation}{A.\arabic{equation}}\setcounter{equation}{0}
In this appendix we calculate the Ricci tensor of the off-diagonal metric \eqref{eq:ometric}\,.
When we use the basis $\phi^i\,(i=1\sim 3)$ for the fiber metric and 
the orthonormal basis {$\hat{e}^0\,,\hat{e}^{\alpha}\,\,(\alpha=1\sim m)$} 
defined by \eqref{eq:o.n.b}\,, the non-zero components are 
\begin{equation}
\begin{aligned}
\widehat{R}_{00}&=-\sum_i \dot{K}^i_i - \sum_{i,j} K^j_i K^i_j - m \cfrac{\ddot{f}}{f}~, \\
\widehat{R}_{ij}&=-\sum_k \dot{K}^k_i b_{kj}- \left(\sum_{\ell} K^\ell_\ell + m \cfrac{\dot{f}}{f} \right)
\sum_k K^k_i b_{kj} \\
&+R^{SO(3)}_{ij} + \cfrac{1}{4 f^4} \sum_{k, \ell, m, n} \sum_{\alpha, \gamma}
b_{ik} b_{j \ell} \O_{mk} F^m_{\alpha \gamma} \O_{n \ell} F^n_{\alpha \gamma}~, \\ 
\widehat{R}_{\alpha \beta}&=-\left(\cfrac{\ddot{f}}{f}+(m-1)\cfrac{\dot{f}^2}{f^2}+\sum_i K^i_i 
\cfrac{\dot{f}}{f} \right)\delta_{\alpha \beta} \\
&+\cfrac{1}{f^2}R_{\alpha \beta}
-\cfrac{1}{2 f^4}\sum_{ijk\ell}\sum_{\gamma}
b_{ij}\O_{ki}F^k_{\alpha \gamma}\O_{\ell j}F^\ell_{\beta \gamma}~, \\
\widehat{R}_{0 i}&=-\sum_{jk}\varepsilon_{ijk} K^k_j~, \\
\widehat{R}_{i \alpha}&=\cfrac{1}{2 f^3}\sum_{jk}\sum_{\beta}b_{ij}\O_{kj}D_\beta F^k_{\alpha \beta}~.
\end{aligned}
\end{equation}
Here $K^j_i=(1/2)\dot{b}_{ik} b^{kj}$ and $R^{SO(3)}_{ij}$ denotes the Ricci tensor on $SO(3)$~;
\begin{equation}
\begin{aligned}
R^{SO(3)}_{ij}&=-\sum_{k \ell} \Gamma^k_{i\ell}\Gamma^\ell_{jk}~, \quad
\Gamma^k_{ij}&=\cfrac{1}{2}\left(\varepsilon_{ijk}+\sum_{\ell m}\varepsilon_{i\ell m}b_{mj}b^{k\ell}
+\sum_{\ell m}\varepsilon_{j\ell m}b_{mi}b^{k\ell}
\right)~.
\end{aligned}
\end{equation}
When we impose the assumption of 4n-dimensional QK manifolds (see section 3)~, 
then $\widehat{R}_{i\alpha}=0$ and the explicit dependence of the group element $\O_{ij}$
disappears from the equations,
\begin{equation}
\begin{aligned}
\sum_{k, \ell, m, n} \sum_{\alpha, \gamma}
b_{ik} b_{j \ell} \O_{mk} F^m_{\alpha \gamma} \O_{n \ell}
F^n_{\alpha \gamma}&=4n \sum_k b_{ik} b_{kj}~, \\
\sum_{ijk\ell}\sum_{\gamma}
b_{ij}\O_{ki}F^k_{\alpha \gamma}\O_{\ell j}F^\ell_{\beta \gamma}&=\sum_i b_{ii}\delta_{\alpha \beta}~,
\end{aligned}
\end{equation}
which yields the $SO(3)$ invariance of the Einstein equation.

\section*{Appendix B}
\renewcommand{\theequation}{B.\arabic{equation}}\setcounter{equation}{0}
Putting $y=b-c$, and then using the Einstein equation \eqref{eq:ein} we obtain
\begin{equation}
\ddot{y}=p\dot{y}+qy~, \label{asymp}
\end{equation}
where the coefficients $p$ and $q$ are given by
\begin{equation} 
\begin{aligned}
	p&=-\frac{\dot{a}}{a}-4n\frac{\dot{f}}{f}~, \\ 
	q&=-\frac{\dot{b}\dot{c}}{bc}
		-\frac{1}{2a^2b^2c^2}(a^2-b^2-c^2)(a^2+ b^2+2bc+c^2) \\
 	&+\frac{n}{f^4}(b^2+bc+c^2)-\Lambda~.
\end{aligned}
\end{equation}
Taking account of the expansion \eqref{ex1}, we approximate the equation 
\eqref{asymp} in the limit $t \rightarrow 0$,
\begin{equation}
	\ddot{y}+\frac{1}{t}\dot{y}-\frac{4N^2}{t^2}y = 0, 
\end{equation}
which has the regular solution $y=t^{2N}$ for a positive integer $N$. 
Thus we conclude that the
expansion takes the form of \eqref{ex3}.

\section*{Appendix C}
\renewcommand{\theequation}{C.\arabic{equation}}\setcounter{equation}{0}
Let us consider $k\neq1$ case, where cone-like singularity appears 
at T-bolt. The angle around T-bolt is $2\pi k$ then the singularity 
is angular deficit for $k<1$ and angular excess for $k>1$. 
In this case, solutions make a family parameterized by $k$. 
A curve on the $(f_0, g_0)$-plane depicted in Fig. 5 shows a family 
of solutions. 

\begin{figure}[h]
  \begin{center}
    \includegraphics[height=7cm]
		{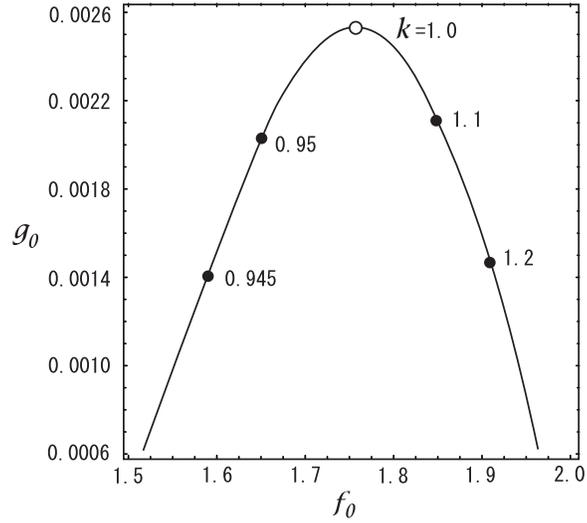}
\parbox{14cm}{
  \caption{\small 
The curve on $(f_0, g_0)$-plane which denotes 
a family of solutions with a cone-like singularity in $n=1$ case. 
The open circle on the curve, critical point $(f_0^*, g_0^*)$, 
is the solution with $k=1$, which is regular at T-bolt.
The left branch with respect to the critical point corresponds to 
the solutions with angular deficit and the right branch does 
angular excess.
}}
  \end{center}
    \label{contour}
\end{figure}


\end{document}